\documentclass[sigconf]{acmart}
\AtBeginDocument{%
  }

\copyrightyear{2026}
\acmYear{2026}
\acmDOI{XXXXXXX.XXXXXXX}
\acmConference[ICSE 2026]{IEEE/ACM International Conference on Software Engineering}{April 12--18, 2026}{Rio de Janeiro, Brazil}

\usepackage{xcolor} 
\usepackage{soul} 
\definecolor{mygray}{gray}{0.85}
\sethlcolor{mygray}

\usepackage{xspace}
\newcommand{\ie}{\emph{i.e.,}\xspace}
\newcommand{\eg}{\emph{e.g.,}\xspace}

\usepackage{pifont}

\begin{document}

\title[Big Questions on Software Architecture]{Big Questions on Software Architecture: \\Report of the ICSE 2026 BoF on Software Architecture}


\author{Davide Taibi}
\orcid{0000-0002-3210-3990}
\affiliation{%
  \institution{University of Southern Denmark}
  \city{Vejle}
  \country{Denmark}}
\email{taibi@imada.sdu.dk}

\author{Patricia Lago}
\orcid{0000-0002-2234-0845}
\affiliation{%
  \institution{Vrije Universiteit Amsterdam}
  \city{Amsterdam}
  \country{The Netherlands}}
\email{p.lago@vu.nl}

\author{Henry Muccini}
\orcid{0000-0001-6365-6515}
\affiliation{%
  \institution{University of L'Aquila}
  \city{L'Aquila}
  \country{Italy}}
\email{henry.muccini@univaq.it}

\renewcommand{\shortauthors}{Taibi et al.}

\begin{abstract}
Software Architecture (SA) is more important than ever, from the need to master the pervasive impact of technology in society and planet earth, to balancing the role of humans versus technology, fle\-xi\-bility versus chaos, economic convenience versus digital autonomy. Despite SA importance, we feel that the SA research community needs to reflect on what are the big questions that deserve scientific research, and what are instead technology trends or hypes.

In this paper we report the results of a community initiative aiming to identify and discuss the \textit{big questions} in SA research. Organised in two sessions hosted in 2026 at the ICSE and ICSA conferences, respectively, this paper reports on the results collected during the first session, and looks forward to complementing them with the results of the second one. \textit{To be continued \dots}

\end{abstract}




\maketitle

\section{Introduction}

Software Architecture (SA) has historically played a central role in supporting the design, evolution, maintainability, and quality of software systems. Over the years, the discipline has continuously evolved to address emerging technological paradigms, including service-oriented architectures, cloud-native systems, microservices, cyber-physical systems, and, more recently, Artificial Intelligence enabled (AI-enabled) and autonomous systems.

At the same time, the rapid pace of technological evolution has increasingly shifted research attention toward short-term trends and technology-driven challenges. While these developments have generated important advances, they have also raised concerns regarding the long-term foundations of the discipline, the sustainability of current research directions, and the enduring scientific questions that should guide future SA research.

In parallel, SA research is facing new challenges related to the growing complexity of modern software systems, the increasing role of AI in software development and decision making, the loss and fragmentation of architecture knowledge, and the need to support continuously evolving distributed systems that operate in highly dynamic environments.

These challenges motivate the need for a meaningful reflection within the Software Engineering and SA communities on the future of the discipline. In particular, there is a need to identify the ``big questions'' that are expected to remain relevant over time and that may contribute to shaping future research agendas, foundational principles, and industry practices.

To stimulate this discussion, we organized the \emph{Software Architecture Birds-of-a-Feather (BoF)} session at the IEEE/ACM International Conference on Software Engineering (ICSE 2026). The session was designed as an open community forum where researchers and practitioners could discuss major open challenges, unresolved problems, and future directions for SA research.

This paper reports the outcomes of the first phase of this initiative. More specifically, we present the study design adopted to collect and discuss big questions in SA, together with the main findings emerging from the ICSE 2026 BoF discussions. The work also introduces the second phase of the initiative, consisting of a working session at the IEEE International Conference on Software Architecture (ICSA 2026), aimed at refining and extending the identified research questions through a collaborative discussion with the international community.

\section{Study Design}
The goal of this study is to identify, discuss, and refine the \emph{big questions} that should guide future SA research. 

With this study, we aim to stimulate a meaningful reflection within the Software Engineering and SA communities on the current challenges, foundational principles, and long-term research directions of the discipline. In particular, we aim to identify research questions that are expected to remain relevant over time and that may contribute to shaping the future of SA research and practice.

To achieve this goal, we designed a community-driven and iterative study composed of two complementary phases involving interactive discussions with researchers and practitioners from the Software Engineering and SA communities.

The study is structured as follows:

\begin{itemize}
    \item \textbf{Phase 1:} ICSE 2026 Birds-of-a-Feather (BoF) Session on Software Architecture
    \item \textbf{Phase 2:} ICSA 2026 Working Session on Big Questions in Software Architecture Research
\end{itemize}

Phase 1 focused on collecting and discussing candidate big questions emerging from the community, while Phase 2 aims at refining, validating, and extending the findings through a collaborative working session.

\subsection{Phase 1: ICSE 2026 BoF on Software Architecture}

The first phase of the study was conducted during the \emph{Software Architecture BoF} session at ICSE 2026\footnote{\url{https://conf.researchr.org/track/icse-2026/icse-2026-software-architecture-birds-of-a-feather-session}}.

The BoF session was designed as an open and interactive forum aimed at encouraging the Software Engineering community to reflect on the future of SA research. The goal was to create a space for discussing fundamental challenges, unresolved problems, and long-term research questions that are independent from technological hypes and rapidly evolving industrial trends.

\paragraph{Call for Contributions and Selection Process}
We invited researchers and practitioners from the Software Engineering community to submit short position papers discussing what they considered to be important open challenges or ``big questions'' in SA research.

Submissions addressed a wide variety of topics, including foundational concepts, emerging paradigms, industrial relevance, education and skills, evaluation methods, sustainability of research directions, and the future role of software architects.

All submissions underwent a thorough peer-review process focused on relevance, clarity, originality, and their potential to stimulate discussion within the community. At the end of the review process, nine submissions were accepted for presentation and discussion during the BoF session.

\paragraph{Session organisation}
To facilitate focused discussions, the accepted papers were organized into three thematic panels, each composed of three contributions. Prior to the session, the moderators analysed the accepted papers and extracted one representative ``big question'' from each contribution. These questions served as the starting point for the panel discussions and audience interaction.

The three panels were organized along the following identified themes:

\begin{itemize}
    \item \textbf{Software Architecture: Skills and Knowledge}
    \item \textbf{Software Architecture: Novelty and Foundations}
    \item \textbf{Software Architecture: Research Agenda and Directions}
\end{itemize}

Each panel included a moderated discussion among the authors followed by an open debate with the audience. The role of the moderators was to stimulate critical discussion, identify common themes across contributions, and encourage participants to challenge assumptions and discuss possible future research directions.

The objective of the session was not to achieve consensus, but rather to surface diverse perspectives, identify recurring concerns, and highlight possible tensions and gaps in current SA research.

\paragraph{Data Collection and Synthesis}
During the BoF session, moderators collected notes regarding the main discussion points, recurring themes, disagreements, and emerging research directions raised by both panellists and audience participants.

Thereafter, the moderators conducted an offline synthesis activity aimed at distilling the main ``big questions'' emerging from the discussions. This synthesis process involved reviewing:
\begin{itemize}
    \item the accepted position papers;
    \item moderator notes collected during the session;
    \item observations and comments raised during the audience discussions.
\end{itemize}
Upon consensus, t he session was audio-recorded to ease the extraction of the information afterwards.

The outcome of this phase is a preliminary set of synthesized big questions representing the main concerns, challenges, and long-term research directions identified by the community during the ICSE BoF session.

\subsection{Phase 2: ICSA 2026 Working Session on Big Questions in Software Architecture Research}

The second phase of the study will be conducted during the 23rd IEEE International Conference on Software Architecture (ICSA 2026) as a dedicated working session entitled \emph{Big Questions in Software Architecture Research}\footnote{\url{https://conf.researchr.org/track/icsa-2026/icsa-2026-working-sessions}}.

The objective of this working session is to further discuss, refine, validate, and extend the big questions identified during Phase~1. Before the session, the synthesized outcomes from the ICSE BoF discussions are shared among the potential participants in order to establish a common baseline and stimulate informed discussion.

\paragraph{Working Session Design.}
The ICSA session is designed as an interactive focus-group-style workshop involving researchers and practitioners with expertise in SA. The session will encourage collaborative reflection on the relevance, implications, and long-term impact of the identified big questions.

The discussion will focus on:
\begin{itemize}
    \item validating the relevance and clarity of the identified questions;
    \item refining and reformulating the questions where necessary;
    \item identifying missing or underrepresented perspectives;
    \item discussing the long-term scientific and industrial implications of the identified challenges; and
    \item identifying possible foundational principles and future research agendas for the SA community.
\end{itemize}

The session will be moderated by the organizers, who will facilitate the discussion and support the identification of recurring themes, divergent viewpoints, and emerging consensus among participants. Notes collected during the session will be analyzed together with the results from Phase~1 in order to produce the final synthesis of the study.

\paragraph{Expected Outcomes.}
The expected outcome of Phase 2 is a consolidated and community-refined set of big questions for SA research. The resulting questions are intended to provide a starting point for future discussions within the community and to contribute to defining long-term research priorities in the field.

The combination of the ICSE BoF and the ICSA working session enables an iterative and community-centred process in which initial ideas are first collected and debated broadly and are then refined through focused discussion and collective reflection.

\section{Results}

At the time of writing, the results reflect the discussion in Phase 1 (the ICSE 2026 BoF session). This generated a broad discussion on the future of SA research, involving both the panellists and the audience. The discussions highlighted several recurring concerns regarding the foundations of the discipline, the role of architects in increasingly AI-driven systems, the sustainability of current research directions, and the long-term challenges faced by the field.


In the following, we report the main findings emerging from each thematic panel. For each big question we provide a short description along with the related open research problems (if present).




\subsection{Software Architecture: Skills and Knowledge}

This panel (panel P1) focused on the evolving skills and knowledge required by software architects in the age of generative AI and agentic systems. 

\textbf{\hl{Big Question P1.1:} Who is the ``good'' architect in the age of AI?}
The discussion emphasized that generative AI can support software architecting by expanding the range of design alternatives, producing preliminary architectural artefacts, and suggesting initial design decisions. However, participants observed that these outputs cannot be considered final architectural decisions without expert human validation.
The ``good'' architect in the age of AI may therefore not only be the one who can make high-quality decisions, but also the one who can determine when human architectural judgment is required.

\medskip
\noindent
\textbf{Open Research Problems}
\begin{itemize}
\item \textbf{Timing of architectural intervention.} There is no established understanding of when architects should step in during AI-assisted architecting processes.

\item \textbf{Human validation of AI-generated decisions.} Generative AI can produce preliminary decisions and artefacts, but experienced architects remain necessary to finalize, validate, and contextualize them.

\item \textbf{Cost--productivity tradeoff in AI-assisted architecting.} Intervening too early or too late may significantly affect the value, cost, and productivity of using generative AI for SA.

\end{itemize}

\textbf{\hl{Big Question P1.2:} Which architect skills are needed (for LLM-based systems)?}
The panel discussed the skills required to architect systems including large language models (LLMs) and agentic components. 
Participants emphasized that architects need strong foundational skills to verify, challenge, and refine the output produced by generative AI. 
In this sense, traditional architect skills remain essential, but they must be applied in a context where part of the design space is explored or generated by AI-enabled systems. At the same time, the discussion highlighted the emergence of {\em new technical skills} (see below). 
The panel also emphasized the importance of {\em soft and reasoning skills} such as decision making and metacognition. 

\medskip
\noindent
\textbf{Open Research Problems}
\begin{itemize}
\item \textbf{Cognitive debt in AI-generated and AI-assisted systems.} LLM-based and agentic systems may become difficult to understand, maintain, and evolve if their architectural structure and rationale are not made explicit.

\item \textbf{Verification of generative AI outputs.} Architects need foundational skills to assess the correctness, feasibility, and architectural consistency of AI-generated artefacts and decisions.

\item \textbf{AI-specific architect skills.} Future architects may need skills in prompt engineering, agentic behaviour, self-evolving systems, and iterative AI-supported architecture generation.

\item \textbf{Decision making and rationale management.} Architects need to connect decisions to their motivation, assumptions, tradeoffs, and consequences.

\item \textbf{Metacognitive skills for software architecture.} Architects need to reason about their own decision processes and about the limits of AI-generated architectural recommendations.
\end{itemize}

\textbf{\hl{Big Question P1.3:} How should architects decide when a Proof of Concept is needed?}
The discussion focused on the role of Proofs of Concept (PoCs) in architecture decision-making. Participants observed that a PoC can support the quantitative analysis of complex decisions, help communicate architectural alternatives, and enable rapid feedback from stakeholders. The panel emphasized that PoCs are particularly useful when architectural decisions are {\em complex}, {\em controversial}, or {\em difficult to assess} only through qualitative reasoning. In such cases, a PoC may help reduce political or non-technical bias and provide evidence for resolving conflicts among software engineers and stakeholders.



\medskip
\noindent
\textbf{Open Research Problems}
\begin{itemize}
\item \textbf{Systematic use of PoCs.} There is no established process for deciding when a PoC is necessary in SA.

\item \textbf{Quantitative evaluation of complex decisions.} PoCs can support evidence-based assessment of architectural alternatives, but their role in decision-making needs to be better defined.


\item \textbf{Integration of PoC results into architecture rationale.} Future research should investigate how PoC outcomes can be connected to architecture decisions and their rationale.
\end{itemize}

Overall, the panel discussions highlighted a common concern regarding the future role of software architects. Generative AI and agentic systems may expand the possibilities of architecture design, but they also increase the need for expert judgment, cognitive control, and explicit rationale management. Participants repeatedly stressed that future architects will need both foundational architect skills and new AI-specific technical skills to guide, validate, and govern architecture decisions in increasingly AI-assisted and self-evolving systems.


\subsection{Software Architecture: Novelty and Foundations}

This panel (panel P2) focused on the long-term sustainability of SA research foundations in the context of increasingly AI-driven and highly distributed systems. The discussion highlighted concerns regarding the availability of representative architecture knowledge, the evolving role of architecture models, and the challenges of maintaining human understanding and control in systems increasingly influenced by AI-generated decisions.

\textbf{\hl{Big Question P2.1:} How to mitigate data sampling bias?}
The discussion highlighted the lack of realistic and representative architecture datasets and benchmarks for SA research. Participants observed that many empirical studies and AI-based approaches rely heavily on open-source software systems that often do not reflect the complexity, scale, constraints, and organisational realities of industrial systems.

The panel emphasised that the community currently lacks datasets and benchmarks comparable to those that significantly accelerated progress in other fields such as AI. This poses the risk of developing approaches that perform well on academic datasets but are not generalisable to industrial environments.

\medskip
\noindent
\textbf{Open Research Problems}
\begin{itemize}
    \item \textbf{Lack of realistic architecture datasets and benchmarks.} Existing datasets are often based on OSS systems and do not adequately represent industrial architectures and their complexity.

    \item \textbf{Limited external validity of research results.} Many approaches are evaluated on simplified or non-representative systems, limiting their applicability in industry contexts.
\end{itemize}

\textbf{\hl{Big Question P2.2:} How to handle human-AI inconsistencies pertaining to Architecture Knowledge (AK)?}
The panel discussed the increasing role of AI systems in supporting architecture reasoning and decision making. Rather than fully automating architecture decisions, participants emphasized the importance of using AI as a support mechanism for argumentation and reasoning augmentation.

The discussion highlighted several interconnected aspects (summarised in the open research problems below). First, participants raised concerns about the progressive loss of human control when systems become increasingly AI-developed. Automatically generated models, code, and decisions may become difficult for architects to interpret, validate, or challenge, especially in large-scale systems.

Second, the panel discussed the growing problem of architecture knowledge loss in large and long-lived systems. In many organisations, architecture decisions and rationale remain undocumented or exist only in the minds of senior architects. As systems evolve, documentation often becomes outdated, while the actual architecture is reflected in deployment artifacts such as Kubernetes manifests, Docker Compose files, Infrastructure-as-Code definitions, and runtime configurations.

Another important aspect concerns the representation of architecture knowledge for AI-enabled systems. Participants observed that traditional architecture modelling languages and representations were originally designed for human interpretation rather than machine reasoning. Existing modelling approaches may therefore be inadequate for representing the complexity of modern distributed and AI-enabled systems.

The discussion also highlighted the need for explicit representation of architecture intent and rationale that can be shared between humans and AI systems. Some participants suggested investing in architectural manifests or machine-readable representations capable of supporting integrated reasoning across software, infrastructure, deployment, resource, and data layers.

\medskip
\noindent
\textbf{Open Research Problems}
\begin{itemize}
    \item \textbf{Loss of human control in AI-assisted architecture design.} Architects may progressively lose the ability to validate or fully understand AI-generated architectural decisions.

    \item \textbf{Loss of architecture knowledge in large and evolving systems.} Architecture rationale and system understanding are often undocumented or distributed across multiple artifacts and stakeholders.

    \item \textbf{Architectural knowledge extraction and reverse engineering.} Future systems require techniques capable of reconstructing architectural knowledge from deployment artifacts and runtime environments.

    \item \textbf{Representation of architecture knowledge for AI systems.} Traditional modelling languages may not be suitable for AI-driven reasoning and modern distributed systems.

    \item \textbf{Human-AI collaboration in software architecture.} AI should support architectural argumentation and reasoning rather than fully replacing human decision making.
\end{itemize}

\textbf{\hl{Big Question P2.3:} How to accurately model future distributed (automotive) systems?}
This discussion focused on the evolving role of SA models in highly distributed and safety-critical environments. Participants questioned whether architecture models should maintain their traditional documentation-oriented role or evolve toward new purposes, such as enabling AI-assisted reasoning, certification, runtime adaptation, and integrated system analysis.

A major concern emerging from the discussion was the certification and trustworthiness of AI-assisted architecture decisions. In safety-critical application domains such as automotive systems, participants questioned how certification authorities could validate architectures influenced by AI-generated decisions or recommendations.

While human architects currently remain responsible for approving architecture decisions, there is currently no established process for certifying collaborative human-AI architecture reasoning.

\medskip
\noindent
\textbf{Open Research Problems}
\begin{itemize}
    \item \textbf{Modelling future distributed and AI-enabled systems.} Existing modelling approaches may not adequately capture the complexity of highly distributed and continuously evolving systems.

    \item \textbf{Evolving role of SA models.} Architecture models may need to support runtime reasoning, AI collaboration, certification, and adaptive analysis rather than only documentation.

    \item \textbf{Certification and trustworthiness of AI-assisted decisions.} There is currently no established process for certifying architectures influenced by AI-generated decisions in safety-critical application domains.
\end{itemize}

Overall, 
participants repeatedly stressed the need for new representations, datasets, modelling approaches, and reasoning techniques capable of supporting increasingly distributed, AI-driven, and continuously evolving systems while preserving human understanding, trustworthiness, and control.

\subsection{Software Architecture: Research Agenda and Directions}
This panel (panel P3) focused on identifying gaps in SA research in the context of Architecture Reconstruction (AR) and its environmental sustainability, digital sovereignty and security, and self-coding.

\textbf{\hl{Big Question P3.1:} How (and how much) can AI technology help with AR and its sustainability?}
The discussion focused on new perspectives offered on still open problems like AR and self-adaptation, thanks to new technology like LLMs and self-coding. It also challenged the \textit{status quo} by discussing the promises of such technology versus the related energy footprint. 

\medskip
\noindent
\textbf{Open Research Problems}
\begin{itemize}
\item \textbf{AR is much harder than code reconstruction (currently being addressed).} Creating an effective AR pipeline is much harder than the current equivalent effort for code reconstruction. AR should require the AI models to understand SA from diagrams, artefacts, code; verify correctness with the help of (experienced) architect; identify points for improvement; ideally recreate some parts of the SA; and finally measure the fit-for-purpose of the ``evolved'' SA. In other words, an effective AR pipeline should reconstruct the architecture, understand how to change it, and apply the change. This open research problem is not new, but AI techniques promise to provide useful support. The participants also observed that addressing this problem is expected to require significant research with sound results in the medium-to-long time horizon.

\item \textbf{Energy consumption should be minimised based on the frequency of different types of tasks, and the \textit{observed} associated benefits.} The energy footprint of LLMs for AR, for example, should be considered as the combination of a \textit{one-time} (stochastic) training phase, followed by a phase that transforms the results into a set of static rules, and \textit{N-times} reuses of these (now deterministic) rules. From this perspective, the footprint of training and rule generation, can be ``mitigated'' by the prospective benefits of reusing deterministic rules. The transparency and fairness of what to measure and when (\ie in which phase) are key open problems.

\item \textbf{Greening AR.} AR is recognised as a cross-cutting, multi-faceted challenge. As such, the participants could identify various interesting sub-problems:
    \begin{itemize}
    \item \textit{A prerequisite (in fact, for any field) is the need for an energy revolution.} As a society, we need to find different ways to produce more (green) energy.
    \item \textit{The economics of AR should be a proxy for sustainability.} For-profit organisations should be provided with economic KPIs as proxies for their environmental counterparts, so that saving money and gaining new markets create the incentives to reduce energy consumption as a by-product.
    \item \textit{Energy efficiency should be embedded in tooling.} That way, architects (and other software professionals) adopt (\ie carve) green practices, and learn them incrementally and seamlessly.
    \end{itemize}
\end{itemize}

\textbf{\hl{Big Question P3.2:} How should architecture treat security and digital sovereignty by design?}
Most of the discussion focused on the SA relevance of digital sovereignty. Participants recognised that digital sovereignty is a big architecture-relevant question, in terms of both how it should be defined and assessed, and what related principles should guide architecture design and decision making, or even compliance to laws and regulations.

\medskip
\noindent
\textbf{Open Research Problems}
\begin{itemize}
\item \textbf{Sovereignty in SA is not a single QA but an overarching multi-disciplinary principle.} Architecture sovereignty can already be assessed via a combination of several target QAs. However, how SA addresses sovereignty principles ``by design'', like data ownership or software/service autonomy, is an open research question. Such a ``big picture'' could be synthesised into, \eg reference architectures or sovereignty tactics, but first these need to be researched, created and matured. In the same vein, the notion of sovereignty needs to be consolidated and operationalised into SA artefacts and tools that ensure compliance with constraints coming from, \eg new laws and regulations (for compliance) or geopolitical concerns (for digital autonomy).
\\
A related foundational research problem is how to keep separated the data (used to train the models; subject to confidentiality) and the resulting reusable models. It was observed that these reusable models should be, like ``protocols'', general-purpose (or domain-agnostic).
\end{itemize}

\textbf{\hl{Big Question P3.3:} How to enable self-coding software systems?}
The panel discussed some interesting research problems pertaining to architectures exploiting self-coding. Participants recognised that self-coding introduces new uncertainty due to the fact that systems (and the data they produce) may change in unplanned and unknown ways; and that SA research provides the right instruments to set system-level constraints to keep such changes under control.

\medskip
\noindent
\textbf{Open Research Problems}
\begin{itemize}
\item \textbf{Enabling self-coding systems is a human-in-the-lead problem.} Context-awareness and self-adaptation are well-explored topics that apply to self-coding systems, too. Self-coding, however, differs from them in the way the systems adapt (\ie change their behaviour by generating and deploying their code autonomously without the \textit{human-in-the-loop}). Think of self-generating code to process data that the human does not even know or understand. One of the new challenges we see, is designing reference architectures to \eg help setting the boundaries (or guardrails) for self-coding systems, or to define novel types of reliability strategies in such extreme uncertainty, which can, and should, be checked by the \textit{human-in-the-lead}.

\item \textbf{One should prioritise what should be self-coded and what should not.} For example, one could argue that self-coding could be more useful for front-end features than back-end services. This because front-ends are often not fitting all-users' requirements-- a weakness that is often under-estimated. With self-coding, however, front-ends could tune autonomously to the user-specific preferences and needs. Differently, back-end features need to comply to common QAs, functional requirements, and possibly contractual and legal constraints. In other words, we foresee that different degrees of autonomy could apply to different types of architecture elements; identifying which ones, would require a classification that is both application- and domain specific.
\end{itemize}

Overall, the panel discussion pointed to both interesting potential benefits (\eg enabling AR with AI techniques) and necessary mitigations (\eg the need for guardrails and the human-in-the-lead).

\subsection{Software Architecture: Other Directions}

At the end of the session, participants were invited to discuss additional research directions that had not been explicitly addressed during the previous panels. The discussion raised broader questions regarding the future of SA research, particularly in relation to the role of AI in software development and the need to preserve the engineering foundations of the discipline.

\textbf{\hl{Big Question P4.1:} How can AI systems reflect organisational values, ethics, and requirements?}
The discussion highlighted the challenge of ensuring that AI-generated artifacts and decisions are aligned with the values, principles, and requirements of the organisations adopting them. Participants questioned how AI systems can become aware of company-specific constraints, ethical considerations, and organisational objectives when generating software artifacts.

Related concerns also emerged during the discussion on self-coding systems. Participants questioned the ethical implications of systems capable of generating software autonomously and emphasized that, regardless of the underlying technology, such systems still need to be engineered, governed, and supervised by humans.

\medskip
\noindent
\textbf{Open Research Problems}
\begin{itemize}
    \item \textbf{Representation of organisational values and requirements.} There is limited understanding of how organisational objectives, constraints, and ethical principles should be represented and communicated to AI systems.

    \item \textbf{Alignment of AI-generated artifacts with organisational expectations.} AI-generated code and architecture decisions should remain consistent with organisational requirements and constraints.

    \item \textbf{Ethics and governance of self-coding systems.} The ethical implications and governance mechanisms of systems capable of generating software autonomously remain largely unexplored.

    \item \textbf{Human supervision of autonomous software generation.} The role of software architects and engineers in supervising and governing self-coding systems remains an open question.
\end{itemize}

\textbf{\hl{Big Question P4.2:} Are we forgetting the foundations of Software Architecture?}
Several participants reflected on the current direction of SA research and questioned whether the growing attention devoted to AI may divert attention from long-standing SA and Software Engineering challenges. The discussion emphasized that several fundamental problems remain open, including modelling, abstraction, system representation, and interoperability.

Participants also questioned whether some current research directions assume that AI technologies will eventually solve problems that are not yet fully understood from an engineering perspective. The discussion highlighted the importance of continuing to investigate foundational SA concepts while exploring new AI-enabled approaches.

Some participants further questioned whether the community is increasingly following technology trends rather than addressing fundamental scientific questions.

\medskip
\noindent
\textbf{Open Research Problems}
\begin{itemize}
    \item \textbf{Foundational SA concepts.} Several long-standing challenges related to modelling, abstraction, system representation, and interoperability remain open.

    \item \textbf{Role of SA foundations in AI-enabled systems.} The relationship between foundational architectural concepts and emerging AI-enabled systems remains unclear.

    \item \textbf{Balance between foundational and technology-driven research.} The community needs to understand how to address emerging technological trends while continuing to investigate fundamental SA challenges.
\end{itemize}

Overall, the discussion highlighted concerns regarding the future direction of SA research. Participants repeatedly stressed the importance of preserving the engineering foundations of the discipline while investigating the opportunities and challenges introduced by AI-enabled and autonomous systems.

\section{Discussion} \label{sec-discussion}

The discussions across the four themes revealed a number of recurring concerns regarding the future of SA research. While the specific topics differ across panels, several common themes emerged.

First, participants repeatedly discussed the growing role of AI in SA activities. Across the discussions, AI was generally viewed as a technology capable of supporting architecture reasoning, decision making, AR, and software development activities. At the same time, participants consistently emphasized the need to preserve human judgment, control, and responsibility. This concern emerged in discussions related to architecture decision making, AK management, self-coding systems, and the certification of AI-assisted architecture decisions.

Second, the discussions highlighted concerns regarding the foundations of the discipline. While many contributions focused on AI-enabled systems, several participants questioned whether the community is dedicating sufficient attention to long-standing SA challenges. Issues related to modelling, abstraction, AK, interoperability, and SA representation were repeatedly mentioned as problems that remain only partially understood despite decades of research.

Third, participants stressed the increasing importance of AK. Multiple discussions highlighted the need to preserve, reconstruct, represent, and communicate AK in systems that are becoming increasingly complex, distributed, and AI-assisted. AK emerged as a central element for supporting both human reasoning and future AI-enabled architecting processes.

Finally, several discussions raised broader concerns regarding governance, ethics, organisational values, and trustworthiness. Participants questioned how future AI-enabled systems can remain aligned with organisational objectives, ethical principles, and regulatory constraints while preserving transparency and accountability.

Taken together, these discussions suggest that future SA research may need to balance two complementary goals. On one hand, the community needs to investigate how AI can support SA activities and enable new classes of software systems. On the other hand, the community needs to continue addressing foundational SA challenges that remain open and that may become even more relevant in increasingly AI-driven software systems.

The results presented in this paper represent the outcome of the first phase of a community initiative. Phase 2 will be conducted during the ICSA 2026 working session on ``Big Questions in Software Architecture Research''. The objective of that session is to further discuss, refine, challenge, and validate the questions identified during the ICSE BoF. In particular, the ICSA session will provide an opportunity to assess whether the identified questions reflect broader community concerns, identify missing perspectives, and consolidate a community-driven research agenda for future SA research.

\subsection{Threats to Validity}

This study is subject to several limitations.

First, the findings are based on discussions involving the participants of the ICSE 2026 Software Architecture BoF session. Consequently, the identified questions reflect the perspectives of the researchers and practitioners who attended the session and may not fully represent the broader SA community. Running the second session at ICSA offers the opportunity to mitigate this limitation by catering the input from other members of the community, too.

Second, the discussions may have been influenced by the accepted position papers. Since the panel discussions originated from the submitted contributions, participants may have naturally focused on topics already represented by the accepted papers, potentially limiting the emergence of alternative perspectives. Again, the second session at ICSA will mitigate this limitation by being open to free inputs from the participants. 

Third, the organisation of the session may have influenced the resulting discussions. The accepted papers were clustered into thematic panels by the session organizers based on their interpretation of the submissions. Different organisers might have grouped the papers differently, leading to alternative panel structures and potentially different discussion dynamics. To partially mitigate this potential bias, the three organisers initially looked at the themes independently, and achieved consensus afterwards.

Finally, the synthesis of the findings involved interpretation by the moderators and organizers. Although the analysis was supported by session recordings and moderator notes, the extraction and consolidation of the identified questions inevitably involved researcher judgment.


\section{Conclusion} \label{sec:conclusion}

This paper reports the outcomes of the first phase of a community initiative aimed at identifying big questions for future SA research. The first phase was conducted as a BoF session at ICSE 2026 and involved the discussion of nine accepted position papers organized into three thematic panels, followed by an open discussion with the audience.

The discussions highlighted several recurring concerns for the future of the discipline. Participants emphasized the growing role of AI in SA, but also the need to preserve human judgment, architecture reasoning, and responsibility. The discussions also stressed the importance of architecture knowledge, including its preservation, reconstruction, representation, and use in increasingly distributed and AI-assisted systems. At the same time, participants questioned whether the community is sufficiently addressing foundational SA problems, including modelling, abstraction, interoperability, system representation, and the relationship between traditional architecture concepts and emerging AI-enabled systems.

The identified questions should not be interpreted as a final or exhaustive research agenda. Rather, they represent a preliminary synthesis of the concerns, perspectives, and open problems that emerged during the ICSE BoF discussion. Their value lies in providing an initial basis for broader community reflection on the long-term directions of SA research.

The next step of this initiative will be the ICSA 2026 working session on ``Big Questions in Software Architecture Research''. This second phase will be used to further discuss, refine, challenge, and extend the questions reported in this paper. The outcome of the two phases is expected to contribute to a community-refined set of research questions that can support future discussions on the foundations, methods, and priorities of meaningful SA research.

\end{document}